\documentclass[twocolumn,showpacs,amsmath,amssymb,aps,superscriptaddress,nofootinbib,longbibliography]{revtex4-1}
\usepackage[utf8]{inputenc}
\usepackage{graphicx}
\usepackage{physics}
\usepackage{bm}
\usepackage{hyperref}
\usepackage{algpseudocode}
\usepackage{algorithm}
\usepackage{subcaption}
\algrenewcommand\textproc{\texttt}

\usepackage{fdsymbol}
\usepackage{tcolorbox}
\usepackage{braket}
\usepackage{amsmath}
\newcommand{\angstrom}{\textup{\AA}}
\usepackage{amssymb}

\usepackage{graphicx}
\usepackage{bbold} 
\usepackage{tikz}
\usepackage{booktabs}

\usepackage{xspace}
\usepackage{xcolor}
\usepackage{dsfont}

\usepackage{etoolbox}

\newcommand{\obj}{\ensuremath{\bar{\mathcal{C}}}}

\newcommand{\CC}{\mathcal{C}}

\newcommand{\II}{\mathbb{I}}

\renewcommand{\epsilon}{\varepsilon}

\newcommand{\vtheta}{\vec{\theta}}

\begin{document}

\title{Performance comparison of optimization methods on variational quantum algorithms}
\begin{abstract}
Variational quantum algorithms (VQAs) offer a promising path toward using near-term quantum hardware for applications in academic and industrial research.
These algorithms aim to find approximate solutions to quantum problems by optimizing a parametrized quantum circuit using a classical optimization algorithm.
A successful VQA requires fast and reliable classical optimization algorithms.
Understanding and optimizing how off-the-shelf optimization methods perform in this context is important for the future of the field.
In this work, we study the performance of four commonly used gradient-free optimization methods: SLSQP, COBYLA, CMA-ES, and SPSA, at finding ground-state energies of a range of small chemistry and material science problems.
We test a telescoping sampling scheme (where the accuracy of the cost-function estimate provided to the optimizer is increased as the optimization converges) on all methods, demonstrating mixed results across our range of optimizers and problems chosen.
We further hyperparameter tune two of the four optimizers (CMA-ES and SPSA) across a large range of models and demonstrate that with appropriate hyperparameter tuning, CMA-ES is competitive with and sometimes outperforms SPSA (which is not observed in the absence of hyperparameter tuning).
Finally, we investigate the ability of an optimizer to beat the `sampling noise floor' given by the sampling noise on each cost-function estimate provided to the optimizer.
Our results demonstrate the necessity for tailoring and hyperparameter-tuning known optimization techniques for inherently-noisy variational quantum algorithms and that the variational landscape that one finds in a VQA is highly problem- and system-dependent.
This provides guidance for future implementations of these algorithms in the experiment.
\end{abstract}

\author{Xavier Bonet-Monroig}
\affiliation{Instituut-Lorentz, Universiteit Leiden, 2300 RA Leiden, The Netherlands}

\author{Hao Wang}
\affiliation{Leiden Institute of Advanced Computer Science, Universiteit Leiden, 2333 CA Leiden, The Netherlands}

\author{Diederick Vermetten}
\affiliation{Leiden Institute of Advanced Computer Science, Universiteit Leiden, 2333 CA Leiden, The Netherlands}

\author{Bruno Senjean}
\affiliation{ICGM, Univ Montpellier, CNRS, ENSCM, 34090 Montpellier, France}
\affiliation{Instituut-Lorentz, Universiteit Leiden, 2300 RA Leiden, The Netherlands}

\author{Charles Moussa}
\affiliation{Leiden Institute of Advanced Computer Science, Universiteit Leiden, 2333 CA Leiden, The Netherlands}

\author{Thomas B\"{a}ck}
\affiliation{Leiden Institute of Advanced Computer Science, Universiteit Leiden, 2333 CA Leiden, The Netherlands}
\author{Vedran Dunjko}
\affiliation{Leiden Institute of Advanced Computer Science, Universiteit Leiden, 2333 CA Leiden, The Netherlands}

\author{Thomas E. O'Brien}
\affiliation{Google Quantum AI, 80636 Munich, Germany}
\affiliation{Instituut-Lorentz, Universiteit Leiden, 2300 RA Leiden, The Netherlands}

\date{\today}
\maketitle

\section{Introduction}
Recently we have witnessed an explosion of quantum computer prototypes accessible to researchers in academic and industrial laboratories.
Existing quantum hardware has already demonstrated the ability to outperform classical computers in specific mathematically contrived tasks~\cite{Arute2019,zhu2021quantum}.
However, it is still unclear whether noisy intermediate-scale quantum (NISQ)~\cite{preskillNISQ} hardware can outperform classical computers on practically useful tasks.
Here, variational quantum algorithms (VQA)~\cite{peruzzo2014variational,mcclean2016theory,cerezoVQA2021} were introduced as a means of preparing classically-hard quantum states by tuning parameters of a quantum circuit to optimize a cost function by utilizing a classical optimizer.

The overall performance of a VQA depends heavily on the performance of the classical optimization algorithm.
Delving into the limitations of these optimization methods for different VQA tasks is critical for future research and industrial applications.
In light of this aim, researchers have proposed new classical optimization algorithms that exploit periodic properties of parametrized quantum circuits~\cite{Nakanish2020SOFF,rotosolve2021}.
Due to the hardness of optimizing VQAs some researchers have focused on using machine learning techniques to help optimize VQAs, e.g., using the so-called surrogate models~\cite{Wilson2021,Sung2020models}, or on simpler optimizers such as the simultaneous perturbation stochastic approximation (SPSA)~\cite{cade2020strategieshubbard}.
The latter performs surprisingly well in a VQA setting, given that it has relatively weak performance in classical benchmarks~\cite{FinckB10c} compared to more complex evolutionary strategies~\cite{AugerBH10h,HansenO01}.
These articles benchmark new optimization techniques relative to standard classical optimizers on a wide variety of systems.
However, in our view, there exists a need for further in-depth analysis of the most common optimization algorithms for VQAs.
One reason is that the existing VQA performance studies only consider testing classical optimizers using default hyperparameters (parameters controlling for the optimizer itself, such as e.g. step size, population size, etc.).
Such a consideration is improper for studying the performance limitation of classical optimizers since, in the classical communities (e.g., Automated Machine Learning), it is well-known that the optimal hyperparameter setting of any optimizer varies drastically across different cost functions, and so does its performance.

In this work, we conduct a systematic hyperparameter tuning of various classical optimization algorithms and compare their empirical performance on a range of problem sizes and at different sampling noise strengths.
We begin by extending the three-stage sampling strategy used in Ref.~\cite{cade2020strategieshubbard} to split the total budget of function evaluations to a suite of four optimizers.
We compare the one- and three-stage sampling on three quantum system, i.e.,H$_4$ square, H$_4$ chain, and $2\times 2$ Hubbard model and find that switching to a three-stage strategy yields mixed results, sometimes improving but sometimes significantly decreasing performance.
We suggest that whether or not multiple stages should be used in optimization should be treated as a hyperparameter for a variational optimizer in future work.

Next, we focus on comparing hyperparameter-tuned SPSA with the state-of-the-art gradient-free optimizer, CMA-ES~\cite{HansenO01}, on a range of seven problems of varying qubit size.
We find that after tuning the hyperparameters of both optimizers, they yield very similar performance.
This is surprising, as previous studies of noisy classical optimization testbeds~\cite{FinckB10c,AugerBH10h} suggest that SPSA is already significantly outperformed by CMA-ES for mid-sized problems.
We thus conjecture that the quantum cost landscape might be substantially different from that of the classical ones, which is worth investigating in the future. 

Finally, we investigate the effect that a noisy evaluation of the cost function can return a value below the true one, breaking the variational principle.
Yet the parameters of this evaluation provide a much worse approximation to the exact solution.
We see that the native ``favorite'' solution of CMA-ES - a robust statistical estimate of the optimal variational parameter - yields the best result in all cases considered, over the lowest energy observed during the optimization trace, and that it is not statistically sound to apply standard error estimates to the latter.

\section{Background}\label{sec:background}
A variational quantum algorithm attempts to find approximate ground states of an \(N\)-qubit quantum system as the output of a circuit \(U(\vec{\theta})\) with tunable parameters \(\vtheta\).
This generates a variational ansatz,
\begin{eqnarray}
    \ket{\Psi(\vec{\theta})} &=& U(\vec{\theta}) \ket{\Phi},
    \label{eq:parametrized_evol}
\end{eqnarray}
where the parameters \(\vec{\theta}\in[0,2\pi]^d\) control the rotations of single and two-qubit gates in a quantum circuit implementation of \(U\) applied to an initial state \(\ket{\Phi}\) (i.e., \(U(\vec{\theta})=U_k(\theta_k)U_{k-1}(\theta_{k-1}) \dots U_0(\theta_0)\ket{\Phi}\)).
During a VQA run, these parameters are tuned to optimize a cost function \(\CC(\vec{\theta})\), which in our case is the expectation value of a Hermitian observable \(\hat{O}\) relative to the state \(|\Psi(\vec{\theta})\rangle\),
\begin{equation}
    \CC(\vec{\theta})=\langle\hat{O}\rangle=\langle\Psi(\vec{\theta})|\hat{O}|\Psi(\vec{\theta})\rangle.
\end{equation}
To measure the expectation value of \(\hat{O}\) without additional quantum circuitry, it is typical to write \(\hat{O}\) as a linear combination of easy-to-measure operators, i.e., Pauli operators \(\hat{P}_i\in\{\II,X,Y,Z\}^{\otimes N}\)
\begin{equation}
    \hat{O}=\sum_ic_i\hat{P}_i\rightarrow\CC(\vtheta) = \langle \hat{O} \rangle = \sum_i c_i \langle \hat{P}_i \rangle.
    \label{eq:pauli_decomp}
\end{equation}
A VQA then passes the estimation of the cost function \(\CC(\vec{\theta})\) to some classical optimization routine to find the values of \(\vec{\theta}\) minimizing \(\CC\).
This optimization loop and the optimizer choices are the focus of this work.

To get an estimate of the expectation value \(\langle\hat{P}_i\rangle\), one prepares and measures the state multiple times in the \(\hat{P}_i\) basis and calculates the mean of the eigenvalues observed.
This approximates the cost function \(\CC\) by an estimator \(\obj\), whose distribution is dependent on the number of repetitions \(M\) used to calculate \(\langle \hat{P}_i \rangle\),
\begin{equation}
    \obj(\vtheta, M) = \sum_i c_i \big[\langle \hat{P}_i \rangle + \epsilon_i(M)\big].
    \label{eq:sampled_cost_func}
\end{equation}
Here, \(\epsilon_i\) is a random variable drawn from a binomial distribution with variance \(\sigma_i^2\sim 1/M\) that is used to simulate the experimental shot or sampling noise.
Assuming that Pauli operators are measured independently, the variance of the estimator \(\obj\) may be propagated directly,
\begin{equation}
    \mathrm{Var}[\obj]=\sum_ic_i^2\sigma_i^2.
\end{equation}
In general, the assumption of independence is violated.
One may measure mutually commuting operators in parallel~\cite{verteletskyi2019measurement,huggins2021efficient,bonetnearly2020,zhao2020,crawford2021,cotler2020quantum}.
Then the resulting measurement has non-zero covariance~\cite{romero2018strategies,Rubin2018}, which should be accounted for.
However, this only introduces a constant factor to the estimation cost and will not significantly impact the relative optimizer performance.
Here, we will use \(M\), defined in Eq.~\eqref{eq:sampled_cost_func}, as the overall cost for the quantum subroutine which takes \(\vec{\theta}\) and \(M\) as inputs, and outputs \(\obj(\vec{\theta},M)\).

To optimize within a VQA, access to \(\CC(\vec{\theta})\) is provided to a classical optimizer, which then minimizes the sampled cost function \(\obj(\vec{\theta},M)\) as a function of the classical parameters \(\vec{\theta}\).
One can additionally provide estimates of gradients \(\grad_\theta \CC\) (or higher order derivatives) in order to perform gradient-based (or Newton-like) optimization.
To avoid a comparison of the runtime of gradient estimation to that of estimating the raw cost function \(\obj(\vec{\theta},M)\), we only compare four gradient-free optimization algorithms.
Moreover, it has been shown that gradient-based optimization strategies suffer given noisy function evaluations with simple noise structures~\cite{DBLP:conf/icml/WuHXHBZ20} (e.g., stationary and isotropic noisy covariance) in the sense that 1) the convergence rate to local optima is hampered~\cite{DBLP:journals/corr/abs-2006-07904} and 2) such simple noise does not help in escaping from local optima~\cite{DBLP:conf/icml/ZhuWYWM19}.
We select the following optimizers (see Appendix~\ref{app:optimizers} for further details):
\begin{enumerate}
    \item SLSQP determines a local search direction by solving the second-order local approximation of the cost function that satisfies the constraints,
    \item COBYLA uses linear approximations of the target and constraint function to optimize a simplex within a trust region of the parameter space,
    \item CMA-ES is a population-based optimization algorithm where the points are drawn from a multivariate Gaussian distribution, whose parameters (covariance matrix and location) are adapted online,
    \item SPSA employs a stochastic perturbation vector to compute an approximate gradient of the cost function simultaneously.
\end{enumerate}
We compare these algorithms across multiple systems of different sizes and the number of parameters considered.

\section{Three-stage sampling comparison}
\begin{figure*}
    \centering
    \includegraphics[width=1.75\columnwidth]{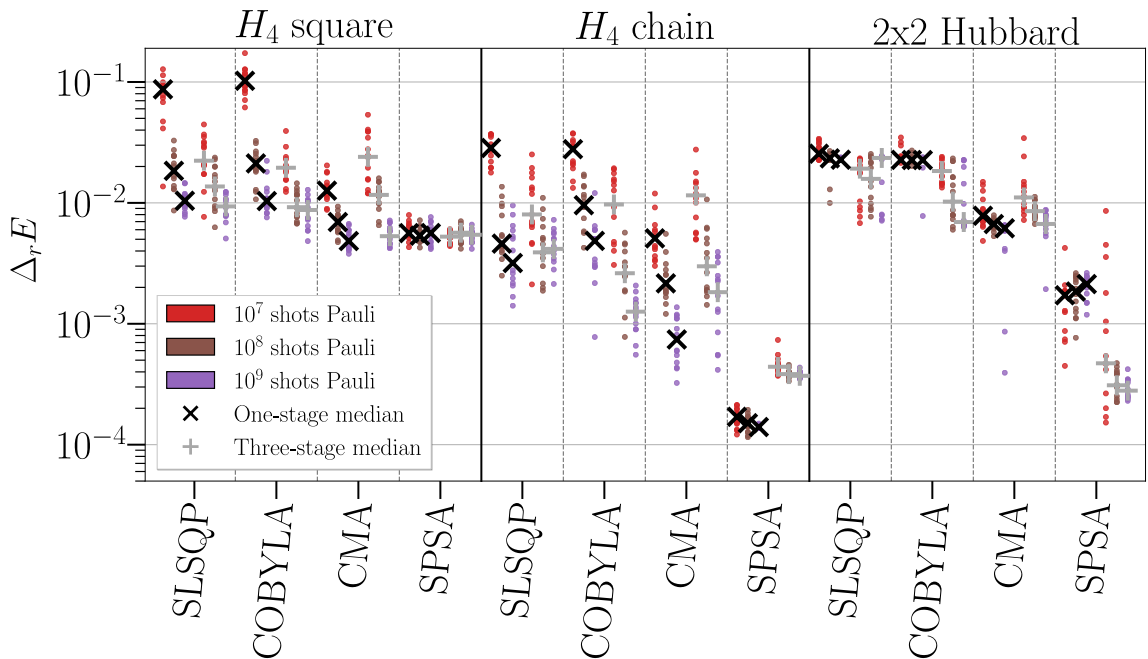}
    \caption{Comparison of optimizers with default hyperparameters for the one-stage (left side) and three-stage (right side) method.
    The final result of the 15 independent runs are shown in dots, with their spread giving an indication of the standard deviation of the sample.
    The colors (red, brown, and purple) indicate the number of total shots used for the optimization.
    Black cross and gray plus depict the median values of 15 independent runs for every problem at every number of shots used.
    }
    \label{fig:unoptimized}
\end{figure*}
Existing state-of-the-art quantum hardware is limited by the stability of the devices, which need to be tuned within time scales of hours up to a day.
This imposes a hard limit on the total number of samples we can measure before the devices changes, of the order of \(\sim\!\!\!10^9\).
This shot budget becomes the limiting factor in VQAs.
One must carefully balance between exploring the parameter space and accurately measuring the cost function.
In this spirit, Cade et al.~\cite{cade2020strategieshubbard} introduced a sampling procedure for SPSA that splits the total shot budget between three stages, resulting in improved performance of VQAs.

Naively, one can fix the total number of shots per Pauli operator and run the optimization until the budget is spent (i.e., if one used 1000 shots per Pauli per function call and allocated a total shot budget of \(10^7\) per Pauli operator, this would allow for a total of 10000 evaluations).
We refer to this approach as one-stage optimization.
Alternatively, one could think of an optimization strategy where the number of shots is increased as the optimization progresses toward better parameters, as introduced by Cade et al. 
We perform a three-stage optimization procedure where the number of samples per Pauli operator increases per phase, reducing the number of total function calls.
In our three-stage optimization, we fix the number of shots per Pauli for each stage (i.e., 100, 1000, 10000 shots for a total budget of \(10^7\)).
The number of function evaluations is then computed from a ratio of 10:3:1.
For every ten function calls at the first stage, we use three function calls in the second stage and one function call in the third stage (i.e., 7150, 2145, and 715 evaluations for a total budget of \(10^7\) shots per Pauli).

We compare the one- and three-stage protocols for the four optimization algorithms previously introduced, aimed at assessing if the three-stage protocol has any evident advantage over the standard sampling strategy.
For this comparison we use the relative energy error,
\begin{equation}\label{eq:figure}
    \Delta_{r}E = \left|\frac{\CC(\vtheta_{\text{opt}}) - E_0}{E_0-c_0}\right|,
\end{equation}
where \(\CC(\vtheta_{\text{opt}})\) is the noiseless cost function evaluated at the optimized parameters \(\vtheta_{\text{opt}}\) obtained from a noisy optimization.
\(E_0\) is the lowest eigenvalue of the problem.
\(c_0\) is the coefficient of the identity operator, which is the largest Hamiltonian term.
This is the relevant figure of merit to capture the performance of the optimizer as it measures the relative error in estimating the traceless part of the Hamiltonian \(H-c_0I\), requiring the quantum computer.
Our numerical experiments are performed under sampling noise with a total shot budget of \(10^7\) (red), \(10^8\) (brown), and \(10^9\) (purple) per Pauli operator.
In the one-stage method, we fix the total number of function evaluations to \(10^4\) and use \(10^3\), \(10^4\), and \(10^5\) shots per Pauli operator at each function call.
In the three-stage procedure, the function calls are also fixed at 7150, 2145, and 715 for all budgets, and the shots per Pauli operator at every stage are \(10^2\), \(10^3\), and \(10^4\); \(10^3\), \(10^4\), and \(10^5\); and \(10^4\), \(10^5\), and \(10^6\), respectively.
The optimization stops when the shot budget is reached.
However, the SLSQP and COBYLA optimization algorithms have a termination criterion that, in most cases, results in exiting early and not utilizing their full shot budget.

We benchmark the algorithms for three different systems on 8 qubits: H$_4$ in a chain and square configuration and the $2\times 2$ Hubbard model.
Throughout this work, we use the Unitary Coupled-Cluster (UCC) variational ansatz for molecules, and the Variational Hamiltonian Ansatz (VHA) proposed in~\cite{cade2020strategieshubbard} for Hubbard models.
Details about the ansatz construction can be found in Appendix~\ref{app:UCCSD} and Appendix~\ref{app:VHA}.
Our results are shown in Fig.~\ref{fig:unoptimized}: for each problem and each optimizer, we show the one-stage (left side with its median indicated with a black cross) and three-stage (right side with its median indicated in a gray plus marker) variants side-by-side.\\
We find that whether a three-stage optimization improves over a one-stage optimization is highly problem- and optimizer-dependent and no trends are statistically significant.
For no single problem is there a consistent gain or loss in median performance across all optimizers.
Moreover, we observe that only the three-stage COBYLA algorithm reports a consistent improvement over its one-stage counterpart across all problems. Given the small number of problems, this is more likely than not to be a fluke.
However, we do see individual instances where going from one-stage to three-stage optimization can cause a significant (5x) performance increase or decrease.
As such, we suggest that though the multi-stage idea of Ref.~\cite{cade2020strategieshubbard} (originally proposed for SPSA and to solve different Hubbard models) can be extended to other problems, it should be treated as a problem-dependent hyperparameter that needs tuning for optimal performance.

Beyond the comparison of one-stage and three-stage optimization strategies, we can make an initial performance comparison between different optimizers in Fig.~\ref{fig:unoptimized}.
We observe that in two out of three problems, SPSA obtains a significantly smaller error than all other optimizers, and in the third it converges faster (though CMA-ES at $10^9$ shots reports a similar performance).
This is at odds with the literature, where CMA-ES in particular is observed to significantly outperform SPSA (see for example a comparison between Ref.~\cite{FinckB10c} and Ref.~\cite{AugerBH10h}).
We conjecture that this is due to the default choice of hyperparameters in our experiments for Fig.~\ref{fig:unoptimized}; optimizing these hyperparameters is known to yield a significant increase in performance~\cite{DoerrYHWSB20,nikolaus_hansen_2019_2594848}.
We hypothesize that, following proper hyperparameter tuning of both methods, the performance of CMA-ES should significantly improve over that of SPSA.

\section{Hyperparameter tuning}
\begin{figure}
    \centering
    \includegraphics[width=\columnwidth]{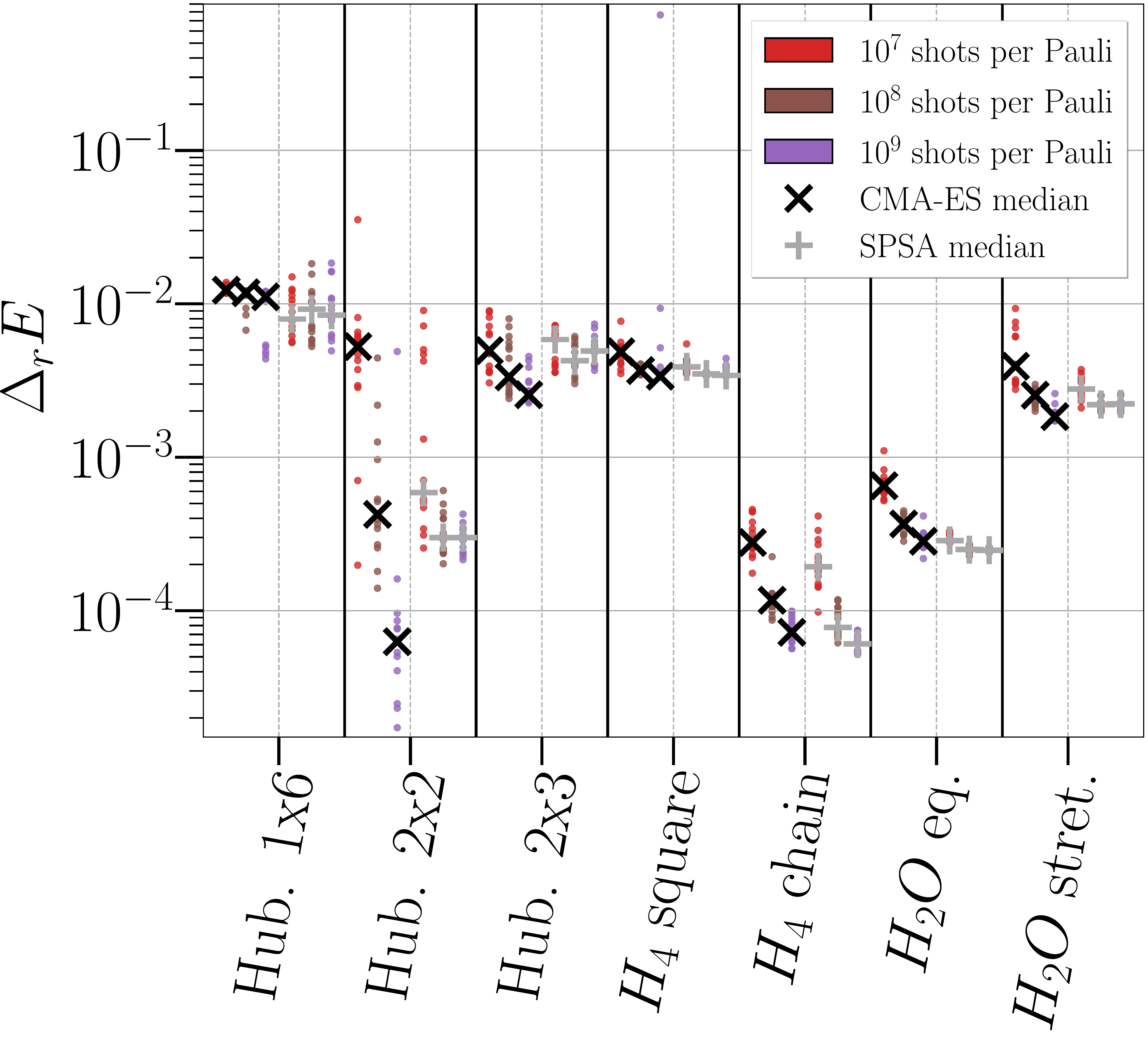}
    \caption{Convergence of hyperparameter-tuned CMA-ES (black cross) and SPSA (gray plus) optimizers across a range of problems and accuracy regimes.
    Dots behind each marker show the 15 independent runs yielding the observed median.
    Each optimizer has their hyperparameters tuned separately for each problem (see App.~\ref{app:hyperparameter-tuning}), treating the one- or three-stage strategy choice as a discrete hyperparameter.
    }
    \label{fig:optimized}
\end{figure}

Most optimization algorithms require a choice of hyperparameters that detail the optimization.
Default hyperparameters are either derived under idealized theoretical assumptions or evaluated from numerical experiments on standard benchmarks.
It is common to tune the hyperparameter of the optimizers when used on a function that has not been previously studied~\cite{DoerrYHWSB20,nikolaus_hansen_2019_2594848}.
Clearly this procedure should be performed similarly for a VQA.

Finding optimal hyperparameters of an optimizer can be costly and generally problem-dependent.
A sub-field of classical optimization has been devoted to automating such hyperparameter tuning.
Here we use the iterated racing for automatic algorithm configuration~\cite{LOPEZIBANEZ201643}, IRACE (see Appendix~\ref{app:hyperparameter-tuning} for a description of the procedure) to tune the SPSA and CMA-ES settings for four molecular systems, H$_4$ square and chain and H$_2$O at equilibrium and stretched geometries (corresponding to weakly- and strongly-correlated regimes, respectively).
Additionally, we perform hyperparameter tuning of CMA-ES for the Hubbard model on three different configurations; $1\times 6$, $2\times 2$, and $2\times 3$.
For SPSA, however, we take the results of ref.~\cite{cade2020strategieshubbard} where the hyperparameters were tuned.
The hyperparameters used for the numerical experiments can be found in Tables I and II in the Appendix~\ref{app:numerical_experiments}

With tuned hyperparameters, we repeat the previous experiments and study a set of additional problems: H$_2$O in equilibrium and stretched geometries (ten qubits), and Hubbard models on $1\times 6$ and $2\times 3$ lattices (twelve qubits).
We see (Fig.~\ref{fig:optimized}) that hyperparameter tuning yields significant improvement of the performance of CMA-ES across all problems.
By contrast, SPSA is only moderately boosted after hyperparameter tuning.
However, the resulting performance between SPSA and CMA-ES is comparable at \(10^9\) total shots.
The only significant difference is a median 5-fold performance gain from CMA-ES over SPSA for the $2\times 2$ Hubbard model.
We observe minor advantages from CMA-ES at $10^9$ shots for the $2\times 3$ Hubbard and H$_2$O in a stretched geometry, and minor gains from SPSA for the $1\times 6$ Hubbard model, H$_4$ chain, and H$_2$O in an equilibrium geometry, but here the deviation lies well within the range of observed final convergence.
This invalidates our hypothesis, as though CMA-ES no longer underperforms compared to SPSA, it does not significantly outperform it either, in contrast to what has been observed in other classical benchmarks~\cite{FinckB10c,AugerBH10h}.
We further do not observe any clear trend with the number of parameters or number of qubits in the system, as though it would be expected to impact the performance of SPSA, as the gradient approximation of SPSA grows worse with the number of parameters.
We suggest that this may be due to our problem sizes being insufficiently large.
Nevertheless, we speculate that the slight overall performance increase for CMA-ES might be an indication that it might be a preferable choice for generic VQA optimization, assuming high-accuracy is required and hyperparameter tuning is affordable.

In addition to the above, we observe that CMA-ES both converges slower than SPSA and typically has a wider spread in final energy error.
The increased spread matches previously reported empirical studies on classical benchmarks~\cite{VermettenW0DB22}, and can be explained as CMA-ES samples points from a larger region than SPSA.
This makes it possible to escape local minima, but this escape cannot be guaranteed, which explains the spread in performance.
The convergence of CMA-ES is also hindered by the need to build a large covariance matrix, which can be observed in call-by-call optimization traces (see Appendix~\ref{app:convergece}).
This implies that it is not immediately possible to gain from `warm-starting' CMA using a good guess of initial parameters (from e.g. a low-cost SPSA run), as we would still need to build the covariance matrix at this point.
We suggest that when sufficiently low accuracy is needed, SPSA may always be preferable to more complex methods in the VQA setting.

\section{The sampling noise floor}
A successful VQA requires the optimization algorithm to return the optimal parameters of \(\obj\).
It is common to take the best-ever measured \(\obj\) as the optimal candidate for the optimization task.

However, in VQAs, the optimization is performed using a proxy cost function \(\obj\) -- a sampled version of the real objective \(\CC\) with variance $\mathrm{Var}[\obj]$.
This may cause us to misidentify a global minimum: given two points $\vtheta_1$ and $\vtheta_2$ such that $\CC(\vtheta_2) > \CC(\vtheta_1) > \CC(\vtheta_2) - a\sqrt{\mathrm{Var}[\obj]}$, we have a chance $\sim\exp(-a)$ to observe $\obj(\vtheta_1) > \obj(\vtheta_2)$.
If we sample repeatedly around the global minimum as we converge and take the best-ever result, this causes two problems.
Firstly, by taking the minimum over so many data points, we sample from a very skewed distribution and can report results with errors far beyond what one might expect from standard $1$- or $2-\sigma$ confidence intervals.
Secondly, the corresponding value of $\vtheta$ might not be a good choice for approximating the ground-state of the problem.
We have a high probability of declaring $\vtheta$ the `best result' even when given access to a noisy estimation $\CC(\vtheta_{g})$ of the cost function at its true minimum as long as $\CC(\vtheta)-\CC(\vtheta_{g})\lesssim \sqrt{\mathrm{Var}[\obj]}$.
This leads us to hypothesise that the values $\vtheta_{\text{best}}$ that gave the best-observed estimates of $\obj$, will have $\CC(\vtheta_{\text{best}}) \gtrsim \CC(\vtheta_{\text{min}})+\sqrt{\mathrm{Var}[\obj]}$.
We call $\CC(\vtheta_{\text{min}})+\sqrt{\mathrm{Var}[\obj]}$ the `sampling noise floor'.

CMA-ES has been designed not to rely on the best-ever function evaluation.
In particular, CMA-ES returns two different candidates; the best-ever measured and a so-called favorite.
The favorite candidate is a statistical estimator of $\vtheta_g$ constructed at the end of the optimization process using all previous measurements.
In principle, such an estimator can average out the sampling noise over the optimization landscape and beat the sampling noise floor.
We note that $\obj(\vtheta_g)$ is never computed during optimization, and so it makes no sense to compare noisy cost function estimates here.

\begin{figure}
    \centering
    \includegraphics[width=\columnwidth]{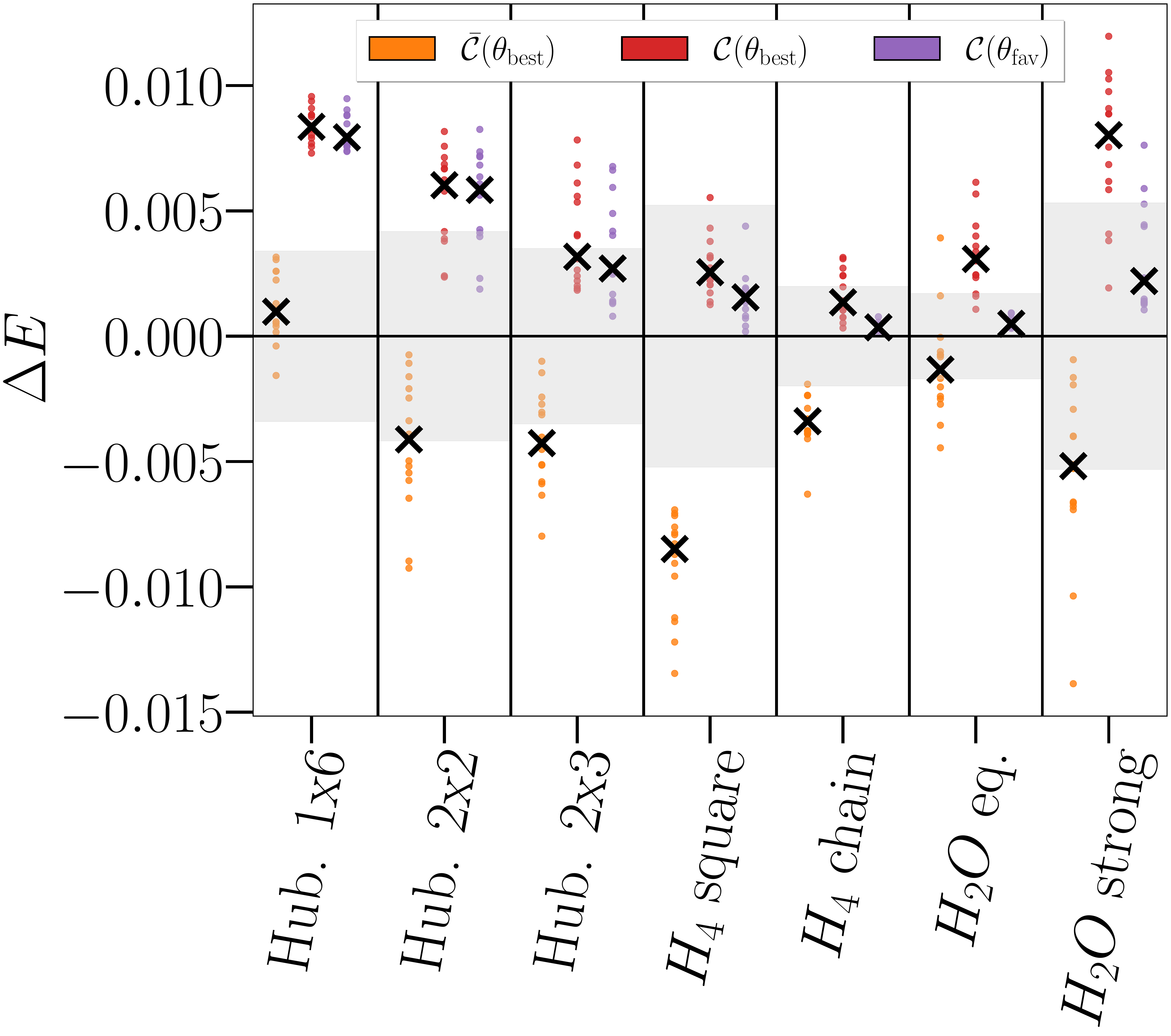}
    \caption{Best-ever versus favorite candidate from CMA-ES under noisy optimization.
    The optimization uses \(10^7\) shots per Pauli over the course of the experiment with individual estimations of \(\obj(\vec{\theta})\) are made using only \(10^4\) shots per Pauli.
    From left to right: (orange) the best-ever measured function evaluation, (red) the best-ever candidate evaluated noiseless, and (purple) the favorite candidate evaluated noiseless. The grey area shows the sampling noise floor for each problem, computed from the best-measured optimization candidate of the \(10^9\) shots scenario.
    The black crosses indicate the sample median of the metric value $\Delta E$ of 15 independent runs of each pair of algorithm and scenario.
    }
    \label{fig:best_fav}
\end{figure}

In Fig.~\ref{fig:best_fav}, we compare the noisy and noiseless evaluation of our cost function at $\theta_{\mathrm{best}}$ to a noiseless evaluation of our cost function at $\theta_{\mathrm{fav}}$ for the optimization traces taken using $10^7$ shots for Fig.~\ref{fig:optimized}.
For each estimated energy we compute the relative energy error \(\Delta E^{(\mathrm{var})} = \frac{\CC(\vtheta_{\text{opt}}) - \CC(\vtheta_{\text{min}})}{\left|\CC(\vtheta_{\text{min}})-c_0\right|}\), where 
$\vtheta_{\text{opt}}=\vtheta_{\text{best}},\vtheta_{\text{fav}}$, and $\CC(\vtheta_{\text{min}})$ is the best found parameters within the variational ansatz from $15$ traces using $10^9$ shots.
(We take this as a proxy for the true variational minimum.)
We observe that the best-ever function evaluation (orange points) is typically below the variational minimum (or our proxy thereof) and 
often below 0 (solid horizontal line), breaking the variational principle.
This error is much larger than the standard deviation $\sqrt{\mathrm{Var}[\obj]}$ (gray shaded region) would suggest, which implies this is no longer a good error bar for $\obj(\vtheta_{\text{best}})$, as predicted.
We further see that, as expected, the true cost function estimates at $\theta_{\mathrm{best}}$ lie around the sampling noise floor.
There are some notable exceptions, namely for the H$_4$ square, H$_4$ chain, and Hubbard $2\times 3$ square.
In all of these cases, the median error is no less than $\sqrt{\mathrm{Var}[\obj]}/2$.
We can explain this by inspecting Fig.~\ref{fig:optimized}: in these cases, the optimizer converges quickly (evidenced by the lack of gain beyond $10^7$ samples), and most function calls are made very close to the true minimum.
We finally observe that in all cases, the median value of \(\CC(\vtheta_{\text{fav}})\) lies below the median value of \(\CC(\vtheta_{\text{best}})\).
Indeed, for certain problems, the difference is significant, and in some cases (H$_4$ chain), the gain over the sampling noise floor can be up to ten-fold.
This suggests that generating a statistical estimator for $\theta_{\mathrm{fav}}$ instead of taking the best value observed is a necessity for an accurate result of any VQAs.

\section{Conclusion}
Variational quantum algorithms have recently prompted significant interest as candidates amenable to near-term hardware.
However, the  performance of these quantum algorithms relies on a classical optimization of a difficult cost function. This task is, in general, intractable to solve optimally.
It is, therefore, important to benchmark the available optimizers for this purpose.
We study the performance of four optimization algorithms under the effect of sampling noise for the task of finding ground-state energies on a range of quantum simulation problems.
We implement the three-stage sampling method of Ref.~\cite{cade2020strategieshubbard} on all four optimizers, and find that the resulting performance improvement is highly problem and optimizer dependent.
This suggests such an adaptation should be treated as a discrete hyperparameter for a VQA optimizer going forward.
We next perform a detailed hyperparameter tuning exercise for the SPSA and CMA-ES optimizers on a larger range of $7$ quantum simulation problems.
Hyperparameter tuning is found to significantly improve CMA-ES, but only to the point where we observe a slight performance improvement over hyperparameter-tuned SPSA.
This suggests CMA-ES would be a preferable optimizer in a high-cost, high-accuracy setting, and SPSA may be otherwise.
We finally demonstrate that using the parameter set that produces the best cost estimate during an optimization trace is a statistically unsound procedure for a VQA with sampling noise.
The error on such an estimate can be further below the VQA minimum than error bars would suggest, and a statistical parameter estimate provided by optimizers such as CMA-ES performs better in terms of the evaluation of the noiseless cost function.
Indeed, the energy error from this statistical parameter estimate can be up to $10$ times lower than the energy error from any single function call (which we term the sampling noise floor).
Extending this study to determine how the relative performance of CMA-ES to SPSA varies as one shifts to larger system sizes is a clear target for future work.

\begin{acknowledgments}
The authors would like to thank Carlo Beenakker for the support on this project, and Jordi Tura, Yaroslav Herasymenko, Stefano Polla, Alicja Dutkiewicz, Saad Yalouz and Eleanor Scerri for fruitful discussions.
CM, TB and VD acknowledge support from TotalEnergies.
VD  acknowledges the support of SURF through the QC4QC project.
This work was in part supported by the Dutch Research Council (NWO/OCW), as part of the Quantum Software Consortium programme (project number 024.003.037).
\end{acknowledgments}

\appendix
\newpage

\section{Details on optimization algorithms}\label{app:optimizers}
In this appendix, we provide a more detailed description of the optimization algorithms used in this work.

\begin{itemize}
    \item Simultaneous perturbation stochastic approximation algorithm (SPSA)~\cite{spall1992,spall1998overview} is designed for noisy evaluations of a cost function, where a stochastic perturbation vector (for instance, a vector whose components are independently sampled from the Rademacher distribution) is used to simultaneously estimate all partial derivatives at given a point.
    Compared to the well-known finite difference method to estimate the gradient, which requires \(2d\) evaluations of the cost function defined over \(\mathbb{R}^d\), the stochastic approximation always consumes two evaluations, hence saving many function evaluations when the search dimension is high.
    However, this algorithm does not follow exactly the gradient direction due to the use of stochastic perturbation.
    \item Constrained Optimization BY Linear Approximations (COBYLA)~\cite{powell_1998} is designed for constrained derivative-free optimization. 
    It employs linear approximations to the objective and constraint functions via a linear interpolation given \(M+1\) points (or simplex).
    These approximations are then optimized within a trust region at each step. 
    \item Sequential Least Squares Quadratic Programming (SLSQP)~\cite{kraft1988software,DieterK94} is an implementation\footnote{We took the implementation from the \texttt{scipy} package, which is based on the original software as described in~\cite{kraft1988software}.} of the more general Sequential Quadratic Programming (SQP) approach~\cite{Nocedal2006} for solving constrained optimization problems. 
    Loosely speaking, in each iteration, SQP proposes a local search direction by solving a sub-problem defined at the current search point in which the nonlinear cost function is replaced by its local second-order approximation, and the constraints are approximated by their affine approximation. When there is no constraint, this method degenerates to Newton's method.
    \item Covariance Matrix Adaptation Evolutionary Strategy (CMA-ES)~\cite{HansenO01} is the state-of-the-art direct search algorithm for the continuous black-box optimization problem, which distinguishes itself from other algorithms in the self-adaptation of its internal variables to the energy landscape. Briefly, this algorithm iteratively draws a number of candidate solutions from a multivariate Gaussian distribution, in which the shape of this distribution (e.g., covariance matrix and location) is adapted online based on the evaluated points in its trajectory.
\end{itemize}

\section{Numerical experiments}\label{app:numerical_experiments}
In this appendix, we describe the numerical experiments used to generate the data for the figures of the article.
The code and data to reproduce these figures can be found in~\cite{bonet2021code}.

To generate the target problems, we use the open-source electronic structure package OpenFermion~\cite{mcclean2020openfermion}.
In addition, we generate the molecular systems with the computational chemistry software Psi4 through the OpenFermion plug-in.
The classical numerical simulations are performed using the open-source quantum circuit simulator package Cirq~\cite{GoogleCirq}.
Regarding the optimization methods, we use the Scipy~\cite{2020SciPy-NMeth} software for COBYLA and SLSQP, PyCMA~\cite{hansen2019pycma} for CMA-ES, and an in-house version of SPSA based on the code in~\cite{spsa_original}.

As described in the main text, we focus on the performance of the optimization methods for VQAs under sampling noise.
In order to include the sampling noise, in our experiments, we compute a noisy expectation value for every Pauli operator in the Hamiltonian with a fixed number of shots, as follows:
\begin{enumerate}
    \item Prepare the ideal quantum state, measure \(\langle P_i \rangle\) and \(p = \frac{1 - \langle P_i \rangle}{2}\),
    \item sample \(\tilde{p} = \mathcal{B}(p, M)\) from a binomial distribution with M shots,
    \item compute a noise expectation value \(\langle \tilde{P}_i \rangle = 1 - 2 \tilde{p}\),
    \item calculate the noisy Hamiltonian expectation value as \(\langle \tilde{H} \rangle = \sum_i c_i \langle \tilde{P_i} \rangle\).
\end{enumerate}
This is a good approximation to the sampling noise generated by measuring the expectation values of Pauli operators in real hardware when the number of shots is large enough.
Moreover, we avoid the bottleneck of preparing and measuring the same state multiple times.

In the Fermi-Hubbard model experiments (see App.~\ref{app:VHA} for further details), we set the parameters of the Hamiltonian to \(t =1.0\) and \(U = 2.0\).
The ansatz circuit for these problems is constructed using the Variational Hamiltonian Ansatze (VHA) with 5 layers for the 1 x 6, 2 layers for the 2 x 2, and 4 layers for the 2 x 3 Hubbard model.
These are the minimum number of layers needed to achieve a ground-state fidelity of \(0.99\) in ref.~\cite{cade2020strategieshubbard}.

For the \(H_4\) in the chain configuration, the first hydrogen atom is located at 0.0 in all coordinates, then every atom is separated in the x-direction by \(1.5 \angstrom\).
In the square configuration, we fix the hydrogen atoms in 2-dimensions.
The positions of the atoms are parametrized by their polar coordinates with \(R = 1.5 \angstrom\) and \(\theta = \frac{\pi}{4}\), and we locate them at \((x,y,0),(x,-y,0),(-x,y,0),(-x,-y,0)\) with \(x=R\cos(\theta)\) and \(y=R\sin(\theta)\).
For the water molecule problems, the \((x,y,z)\)-coordinates of the atoms are given in Table~\ref{tab:waterconfigurations}.
Additionally, in both of the problems, we reduce the active space by freezing the lowest two lowest orbitals, thus reducing the problem from 14 qubits to 10 qubits (or from 7 to 5 spin-orbitals).
As a trial state to approximate the ground-state of the molecular systems, we use the so-called Unitary Couple-Cluster ansatze.
A detailed description of how we construct the UCC ansatze can be found in a App.~\ref{app:UCCSD}.

Finally, the total number of parameters to be optimized for each target problem can be found in table~\ref{tab:numberparams}.
\begin{table}
\begin{center}
\begin{tabular}{||c|c|c||}
    \hline
    \(H_2 O\) & Equilibrium & Stretched \\
    \hline
    \(O\) & (0.0,  0.0, 0.1173) & (0.0, 0.0, 0.0)\\
    \hline
    \(H\) & (0.0,  0.7572, -0.4692) & (0.0, 1.8186, 1.4081)\\
    \hline
    \(H\) & (0.0, -0.7572, -0.4692) & (0.0, -1.8186, 1.4081) \\
    \hline
\end{tabular}
\end{center}
\caption{\label{tab:waterconfigurations} Table describing the configurations of the atoms for the two water molecule problems used in this work.}
\end{table}

\begin{table}
\begin{center}
\begin{tabular}{||c|c||}
\hline
System & \# Parameters\\
\hline
\(H_4\) chain &  14 \\
\hline
\(H_4\) square & 10 \\
\hline
\(H_2 O\) eq. & 26 \\
\hline
\(H_2 O\) stret. & 26\\
\hline
Hub. 1 x 6 & 15\\
\hline
Hub. 2 x 2 & 6 \\
\hline
Hub. 2 x 3 &  16\\
\hline
\end{tabular}
\end{center}
\caption{\label{tab:numberparams} Number of parameters of the ansatze for each target problem.}
\end{table}

\section{Optimization algorithms hyperparameters} \label{app:hyperparameter-tuning}
Prior to applying the aforementioned optimizers on VQAs, we also optimize the hyperparameters of those optimizers. Such an extra tuning task aims at bringing up the performance of each optimizer to the maximum, hence facilitating a fair comparison on each problem.
To achieve this task efficiently, we utilize the well-known IRACE algorithm f for the hyperparameter tuning.
Irace has been extensively applied in automated machine learning research for configuring machine learning models/optimizers~\cite{NobelVWDB21,VermettenWBD20}.

Built upon a so-called iterated racing procedure, this algorithm employs a statistical test (usually the Wilcoxon ranked-sum test) to obtain a robust (with respect to the sampling noise in measured energy values) ranking of hyperparameter settings, thereby serving as a suitable choice for our task.
The tuning process with IRACE initiates by fixing the subset of optimizer hyperparameters to be modified, including bounds and potential constraints.
Then, a `race' between randomly sampled values begins.
These configurations are evaluated a fixed number of times, and the less favourable configurations are disregarded based on a statistical test.
The configurations that survived are then raced again until the evaluation budget is depleted or the number of configurations is below a threshold.
Next, IRACE updates the candidate generation model based on the survival configurations and generates a set of new configurations to race against the elites.
The racing procedure is repeated until the total budget is depleted.
The surviving configurations are returned as the optimal configurations of the algorithm.
The final hyperparameters used for the experiments are the average of the survivors are shown in Table~\ref{tab:SPSA-configurations} and~\ref{tab:CMA-configurations}.

In detail, the hyperparameters we tuned are as follows:
\begin{itemize}
    \item For SPSA, we use the following ranges for each hyperparameter: \(a\in [0.01, 2]\), \(\alpha \in [0, 1]\), \(c\in [0.01, 2]\), and \(\gamma\in [0, 1/6]\).
    \item For CMA-ES, we use the following ones: \(\text{population size } \in [30, 130]\), \(\text{c mean} \in [0, 1]\), \(\mu \in [0, 0.5]\), \(\text{Damp. factor} \in [0, 1]\), and \(\sigma_0 \in [0.25, 1.1]\).
\end{itemize}
For running the IRACE algorithm, we allocated 500 evaluations of the hyperparameters as the total budget, as well as a maximum total running time of seven days, and used two evaluations of each hyperparameter at the beginning of each race. Also, we used the F-test to eliminate worse configurations in the racing procedure. The finally suggested configurations in Table~\ref{tab:SPSA-configurations} and~\ref{tab:CMA-configurations} are the best elites from four independent runs of IRACE.

As for COBYLA and SLSQP, we took their default hyperparameter settings, i.e., \(\rho_{\text{initial}} = 0.1\) and Tolerance\(=10^{-8}\) for COBYLA and \(\epsilon=0.055\) and Tolerance\(=10^{-8}\). 

\begin{table}
\begin{center}
\begin{tabular}{||c||c|c|c|c||}
\hline
\textbf{SPSA} & a & \(\alpha\) & c & \(\gamma\) \\
\hline
default & 0.15 & 0.602 & 0.2 & 0.101 \\
\hline
\(H_4\) chain & 1.556 & 0.809 & 0.106 & 0.097 \\
\hline
\(H_4\) square & 0.867 & 0.593 & 0.133 & 0.113 \\
\hline
\(H_2 O\) eq. & 0.103 & 0.878 & 0.149 & 0.131 \\
\hline
\(H_2 O\) stret. & 0.660 & 0.743 & 0.253 & 0.108 \\
\hline
Hubbard & 0.15 & 0.602 & 0.2 & 0.101 \\
\hline
\end{tabular}
\end{center}
\caption{\label{tab:SPSA-configurations}List of values for SPSA hyperparameters used in Fig.~\ref{fig:unoptimized} (default) and Fig.~\ref{fig:optimized} of the main text after tuning using IRACE.
We perform hyperparameter optimization only with \(H_4\) chain and \(H_2 O\) equilibrium and use the same values for the respective square and stretched configurations.
For the Hubbard models, we take the default values as ref.~\cite{cade2020strategieshubbard} suggests their optimality.
}
\end{table}

\begin{table}
\begin{center}
\resizebox{\columnwidth}{!}{
\begin{tabular}{||c||c|c|c|c|c||}
\hline
\textbf{CMA-ES}\footnote{The rest of hyperparameters are set to their default values} & \(\sigma_0\) & Population & \(\mu\) & c mean & Damp. Factor \\
\hline
default & 0.15 & \small\(\lceil 4+3\log(m) \rceil\) & 0.5 & 1.0 & 1.0 \\
\hline
\(H_4\) chain & 0.20 & 149 & 0.383 & 0.293 & 0.665  \\
\hline
\(H_4\) square & 0.309 & 99 & 0.409 & 0.561 & 0.852 \\
\hline
\(H_2 O\) eq. & 0.344 & 99 & 0.460 & 0.192 & 0.770 \\
\hline
\(H_2 O\) stret.  & 0.310 & 104 & 0.380 & 0.802 & 0.819 \\
\hline
Hub. 1 x 6 & 0.9131 & 51 & 0.3814 & 0.3614 & 0.6006 \\
\hline
Hub. 2 x 2 & 0.8561 & 113 & 0.2741 & 0.6317 & 0.6771 \\
\hline
Hub. 2 x 3 & 0.897 & 128 & 0.1898 & 0.988 & 0.8391 \\
\hline
\end{tabular}
}
\end{center}
\caption{\label{tab:CMA-configurations}List of values for CMA-ES hyperparameters used in Fig.~\ref{fig:unoptimized} (default) and Fig.~\ref{fig:optimized} of the main text after tuning using IRACE. Here, \(m\) indicates the number of free parameters of the ansatz in each problem.}
\end{table}

\section{Unitary Coupled-Cluster ansatz based on coupled-cluster amplitudes}\label{app:UCCSD}
Several classes of systems remain challenging to solve, even for the coupled cluster methods, which are considered the golden standard in quantum chemistry.
Those systems are usually plagued by ``quasidegeneracy'', meaning that the wavefunction cannot be decomposed into a single leading component.
This leads to an important deterioration of methods relying on the single determinant assumption (also said to be mono-reference)~\cite{bulik2015can}.
This issue can be partially solved by developing multi-reference coupled cluster approaches (see Refs.~\cite{bartlett2007coupled,lyakh2012multireference} for a review).
Owing to the recent developments of quantum algorithms in the NISQ-era,
there has been a renewed interest in the unitary formulation of the Coupled-Cluster (UCC), which is naturally suited for quantum computation and naturally extendable to generate multi-reference wavefunctions~\cite{peruzzo2014variational,evangelista2019exact}, while being intractable on classical computers~\cite{romero2018strategies}.
Several formulations of UCC have been investigated to go beyond the standard Unitary Coupled-Cluster Singles and Doubles (UCCSD) method where only fermionic excitations from occupied to virtual orbitals (with respect to the reference determinant, usually the Hartree--Fock one) are considered~\cite{romero2018strategies,greene2021generalized,lee2018generalized,mizukami2020orbital,sokolov2020quantum,grimsley2019adaptive}.
However, the number of operators (and thus the number of parameters) can rapidly become problematic if implemented naively.
A powerful approach is provided by the Adaptive Derivative-Assembled Pseudo-Trotter (ADAPT) types of ansatz~\cite{grimsley2019adaptive,claudino2020benchmarking,Yordanov2021,gomes2021adaptive,liu2021efficient,tang2021qubit,zhang2021mutual}, 
which allows us to adaptively increase the number of operators in the ansatz one by one until reaching a given accuracy.
In this work, we employ a different strategy by taking advantage of the amplitudes extracted from the traditional coupled cluster method performed on a classical computer.
In the coupled cluster, the exponential ansatz reads as follows
\begin{eqnarray}
\ket{\Psi(\vec{t})} = e^{\hat{\mathcal{T}}}\ket{\Phi_0},
\end{eqnarray}
where \(\ket{\Phi_0}\) denotes the reference determinant (like the Hartree--Fock wavefunction) and
\begin{eqnarray}
\hat{\mathcal{T}} = \sum_{i=1}^\eta \hat{\mathcal{T}}_i = \sum_\mu t_\mu \hat{\tau}_\mu
\end{eqnarray}
(\(\eta\) denotes the total number of electrons)
is usually truncated to singles and doubles only:
\begin{eqnarray}
\hat{\mathcal{T}}_1 &=& \mathop{\sum_{i \in \rm occ}}_{a \in \rm virt} t^i_a \hat{a}_a^\dagger a_i,\\
\hat{\mathcal{T}}_2 &=& \mathop{\sum_{i>j \in \rm occ}}_{a>b \in \rm virt} t^{ij}_{ab} \hat{a}_a^\dagger \hat{a}^\dagger_b  a_i a_j.
\nonumber
\end{eqnarray}
One could think of determining the CC amplitudes \(\mathbf{t}\) variationally, but this is not convenient in practice because the Baker--Campbell--Hausdorff (BCH) expansion cannot be used (because \(\hat{\mathcal{T}}^\dagger \neq - \hat{\mathcal{T}}\)).
Tractable implementations rely on a non-variational optimization using the ``Linked'' formulation:
\begin{eqnarray}\label{eq:CC}
e^{-\hat{\mathcal{T}}} \hat{H} e^{\hat{\mathcal{T}}} \ket{\Phi_0} = E(\vec{t})\ket{\Phi_0}.
\end{eqnarray}
The amplitudes are then determined by solving a set of non-linear equations defined by projecting Eq.~(\ref{eq:CC}) against a set of excited configurations \(\lbrace \ket{\mu} \rbrace\) (configurations obtained from the excitation operators in \(\hat{\mathcal{T}}\)):
\begin{eqnarray}
\bra{\mu} e^{-\hat{\mathcal{T}}} \hat{H} e^{\hat{\mathcal{T}}} \ket{\Phi_0} = 0,
\end{eqnarray}
for which the BCH expansion can be used, as it can be naturally truncated to the fourth order.

In this work, we computed the coupled cluster amplitudes of all our molecular systems (\(H_4\) chain, \(H_4\) square, and \(H_2 O\)) and defined our UCC ansatz according to these amplitudes.
Instead of implementing UCC naively by considering all possible excitations, we only keep the excitation operators for which the corresponding CC amplitude is non-zero.
This reduces already the total number of operators (and thus the total number of parameters) significantly.
In practice, we use the trotterized-UCC ansatz,
\begin{eqnarray}
\ket{\Psi(\vec{\theta})} = \prod_{\mu} e^{\theta_\mu (\hat{\tau}_\mu- \hat{\tau}_\mu^\dagger)} \ket{\Phi_0}.
\end{eqnarray}
This trotterized form is an approximation (though it may be mitigated by the classical optimization~\cite{grimsley2019trotterized}) which depends on the ordering of the operators.
We decided to order the operators with respect to the value of the CC amplitudes in descending order, meaning that the first operator to be applied to the reference state within the UCC ansatz will be the operator with the highest associated CC amplitude.
We figured out that the operators in our ansatz were also the ones picked by the ADAPT-VQE ansatz~\cite{grimsley2019adaptive}, although the ordering might not different.
However, ADAPT-VQE can add new operators (or select and repeat an already present operator) to reach a higher accuracy.
To avoid performing the (somewhat costly) first ADAPT-VQE steps, one could think about using our strategy first and then applying ADAPT-VQE for a few more steps to increase the pool of operators slightly.
Note that a stochastic classical UCC can also be employed as a pre-processing step to determine the important excitation operators of the UCC ansatz, as shown in the recent work of Filip {\it et al.}~\cite{filip2020stochastic}.

In our numerical experiments, the initial state is always the Hartree--Fock state corresponding to the number of electrons in the system.
The parameters of the circuit are initialized at 0.0.

\section{Variational Hamiltonian ansatz for the Hubbard model}\label{app:VHA}
In this section, we provide the details on the variational Hamiltonian ansatze (VHA) used for the Hubbard model problems.

The Fermi-Hubbard Hamiltonian describes the behaviour of fermions on a lattice of \(n_x\) x \(n_y\) sites.
Fermions can hop to the nearest neighbouring sites with some strength \(t\), and observe a repulsion or Coulomb term of strength \(U\) to move to the same site with the same spin,
\begin{align*}
    H_{\text{Hubbard}} = H_{\text{t}} + H_{\text{U}} = \\
    -t \sum_{(i,j), \sigma} \left( a^{\dagger}_{i\sigma}a_{j\sigma} + a^{\dagger}_{j\sigma}a_{i\sigma} \right)  + U \sum_{i} n_{i\uparrow}n_{i\downarrow}.
\end{align*}
One can further split the hopping term with respect to the vertical and horizontal hopping terms \(H_{\text{t}} = H_{v} + H_{h}\).

The VHA was introduced in ref.~\cite{wecker2015progress} as a means of constructing parametrized quantum states motivated by time-evolution by Troterrization for the Hubbard model.
However, in our numerical experiments, we use the VHA introduced by Cade et al.~\cite{cade2020strategieshubbard} where the horizontal and vertical terms can be implemented in parallel (see Eq.~[2] in the previous reference).
The parametrized quantum state is constructed as
\begin{align}
    \ket{\Psi(\vec{\theta})} = U(\vec{\theta}) \ket{\Phi} =
    \Pi_{l=1}^{L} e^{i \theta_{v_2,l} H_{v_2}} e^{i \theta_{h_2,l} H_{h_2}} \\
    e^{i \theta_{v_1,l} H_{v_1}} e^{i \theta_{h_1,l} H_{h_1}} e^{i \theta_{U,l} H_U}.
\end{align}
The initial state \(\ket{\Phi}\) is the Gaussian state of the non-interacting part of the Hamiltonian, and the parameters of the circuit are set to 0.0.

The Fermi-Hubbard Hamiltonian is generated with the open-source package OpenFermion~\cite{mcclean2020openfermion}.

\section{Optimization convergence traces}\label{app:convergece}
\begin{figure}
    \centering
    \begin{subfigure}{0.5\textwidth}
    \centering
    \includegraphics[width=\textwidth]{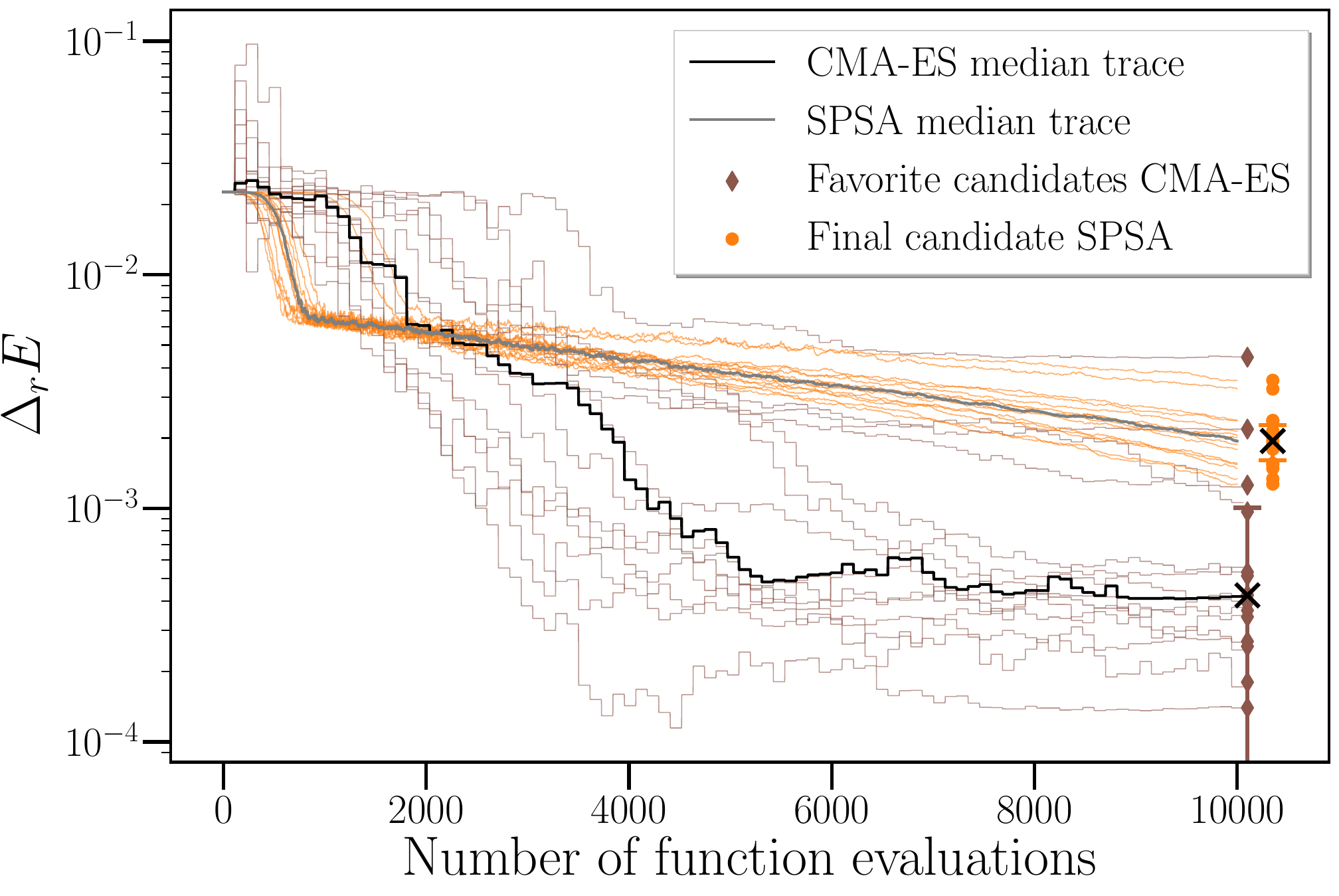}
    \label{fig:appendix_2x2_trace_onestage}
    \end{subfigure}
    
    \begin{subfigure}{0.5\textwidth}
    \centering
    \includegraphics[width=\textwidth]{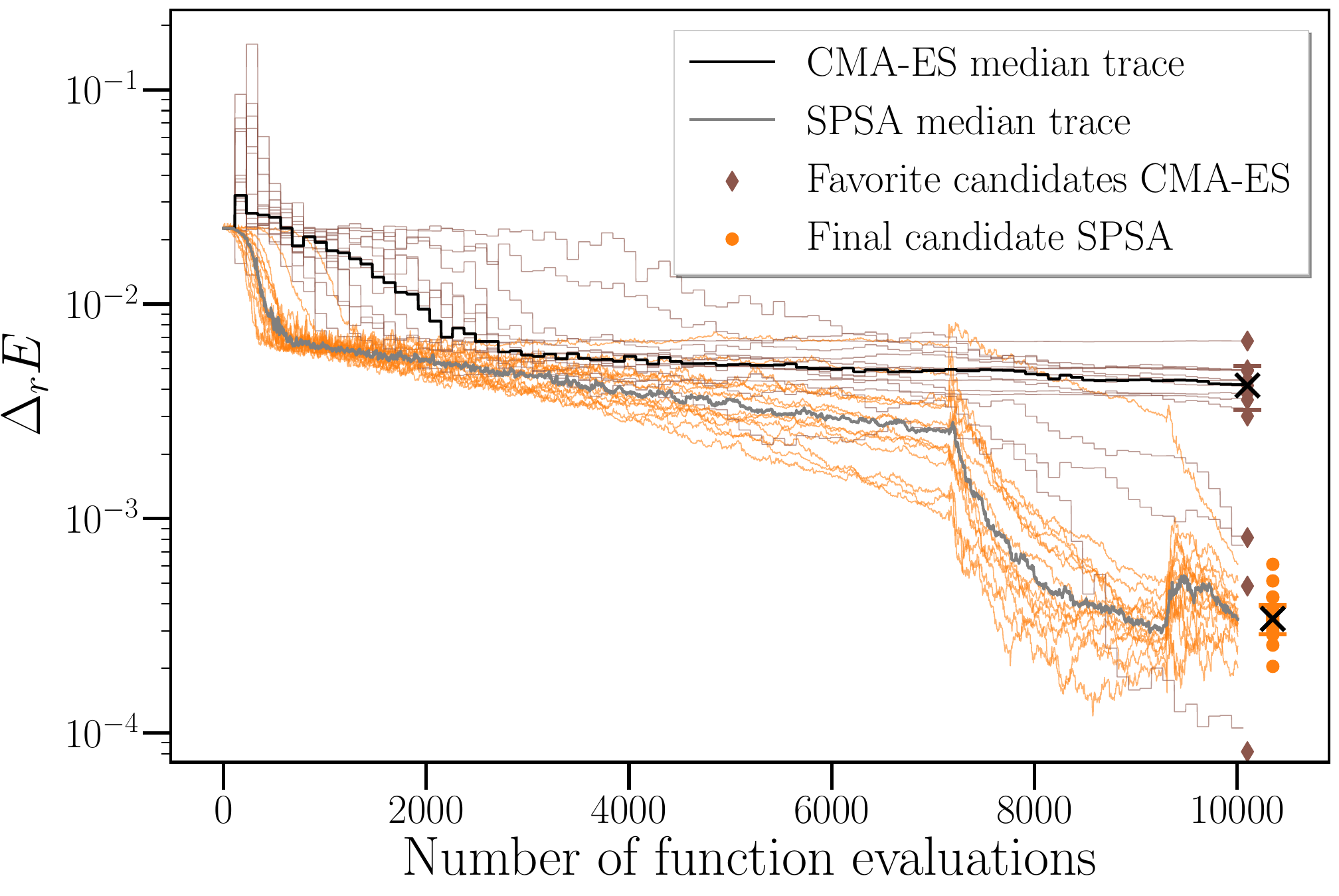}
    \label{fig:appendix_2x2_trace}
    \end{subfigure}

    \caption{Optimization traces of SPSA (orange and grey) and CMA-ES (blue and black) on the 2 x 2 Hubbard model with \(10^8\) total shots.
    The thin lines depict an independent optimization trace, with the thick line on top being the median of the 15 independent runs.
    Top and bottom panel show the one- and three-stage sampling respectively.
    }
    \label{fig:appendix_2x2_trace_both}
\end{figure}

\begin{figure}
    \centering
    \begin{subfigure}{0.5\textwidth}
    \centering
    \includegraphics[width=\textwidth]{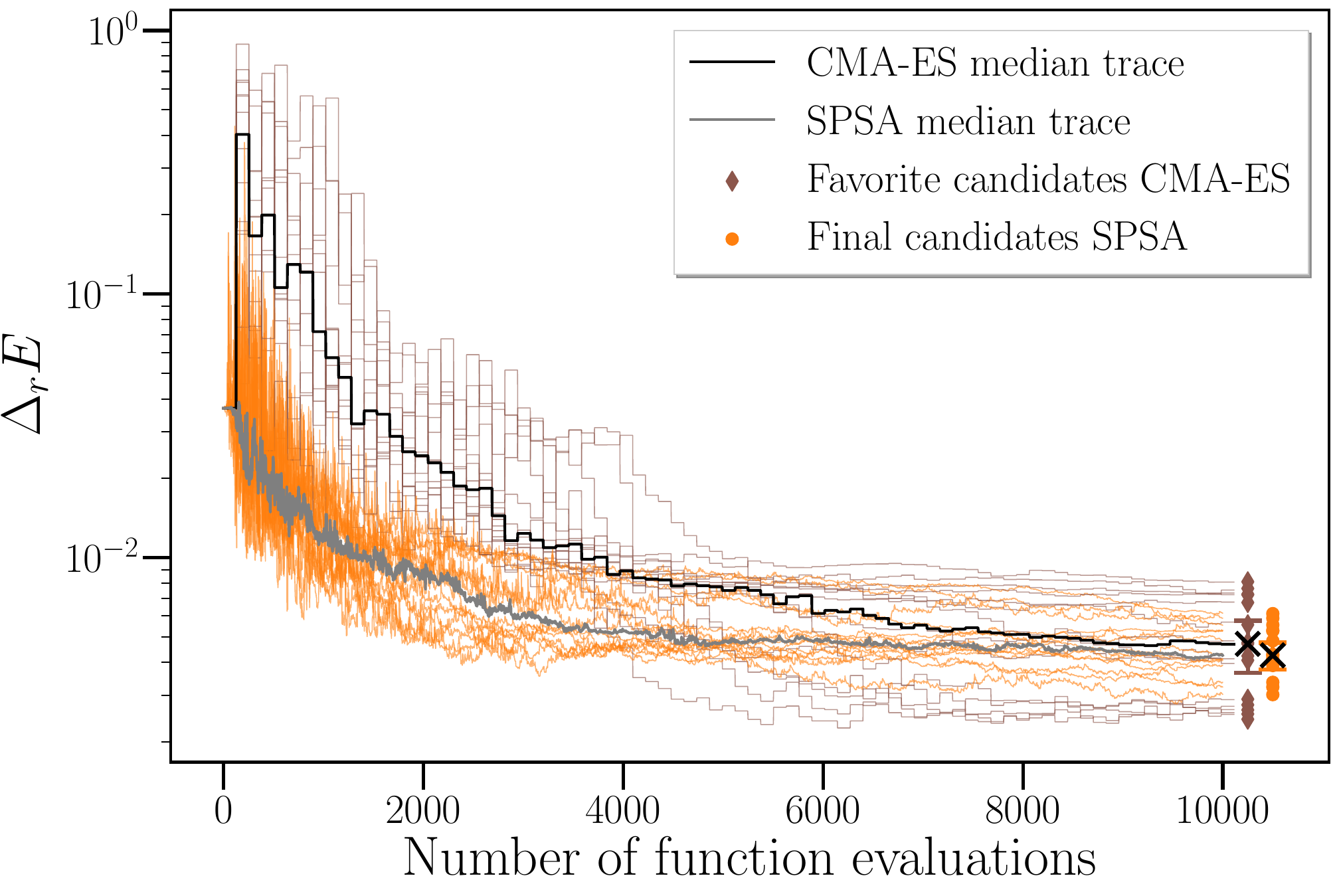}
    \label{fig:appendix_2x3_trace_onestage}
    \end{subfigure}
    
    \begin{subfigure}{0.5\textwidth}
    \centering
    \includegraphics[width=\textwidth]{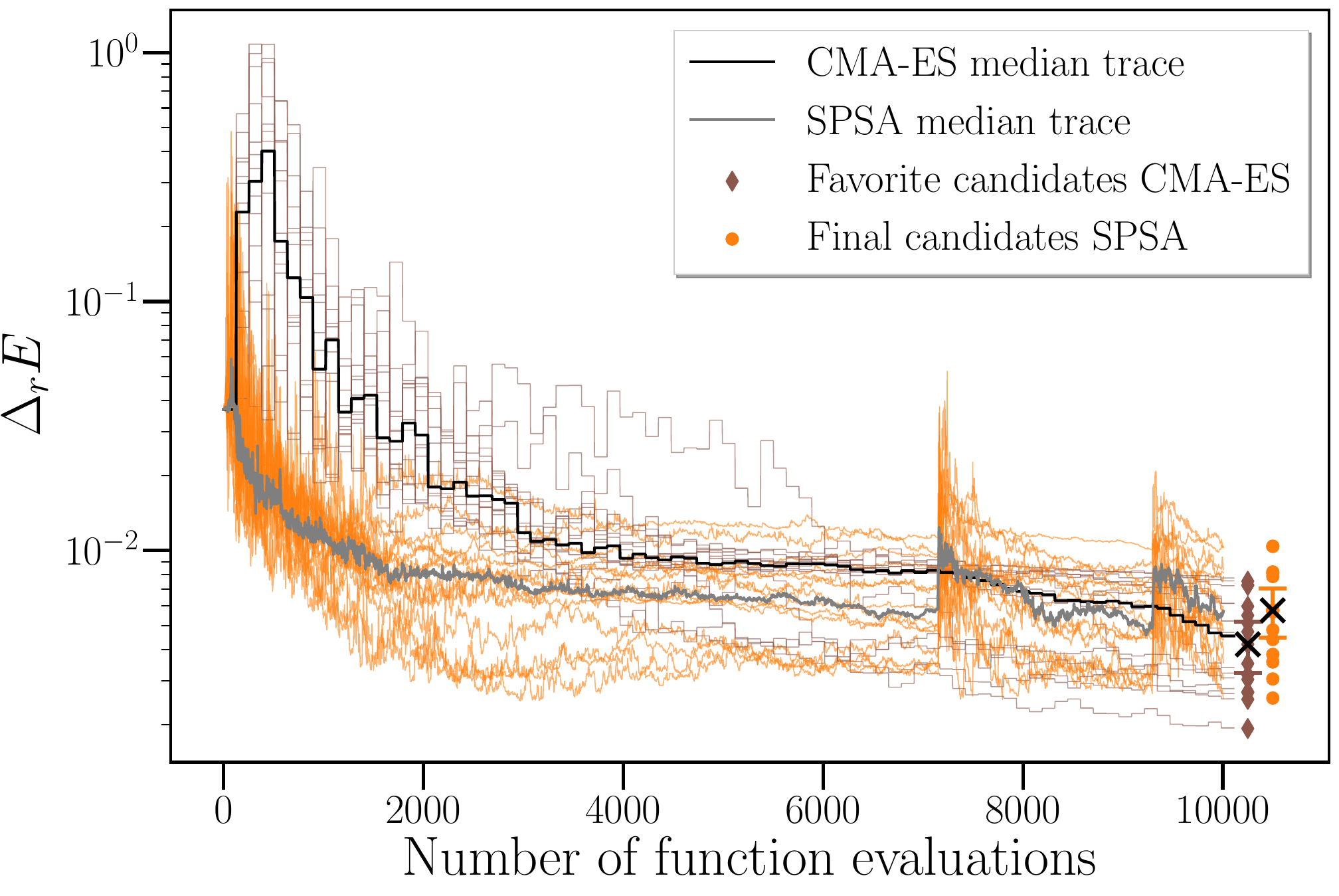}
    \label{fig:appendix_2x3_trace}
    \end{subfigure}

    \caption{Optimization traces of SPSA (orange and grey) and CMA-ES (blue and black) on the 2 x 3 Hubbard model with \(10^8\) total shots.
    The thin lines depict an independent optimization trace, with the thick line on top being the median of the 15 independent runs.
    Top and bottom panel show the one- and three-stage sampling respectively.
    }
    \label{fig:appendix_2x3_trace_both}
\end{figure}

\begin{figure}
    \centering
    \begin{subfigure}{0.5\textwidth}
    \centering
    \includegraphics[width=\textwidth]{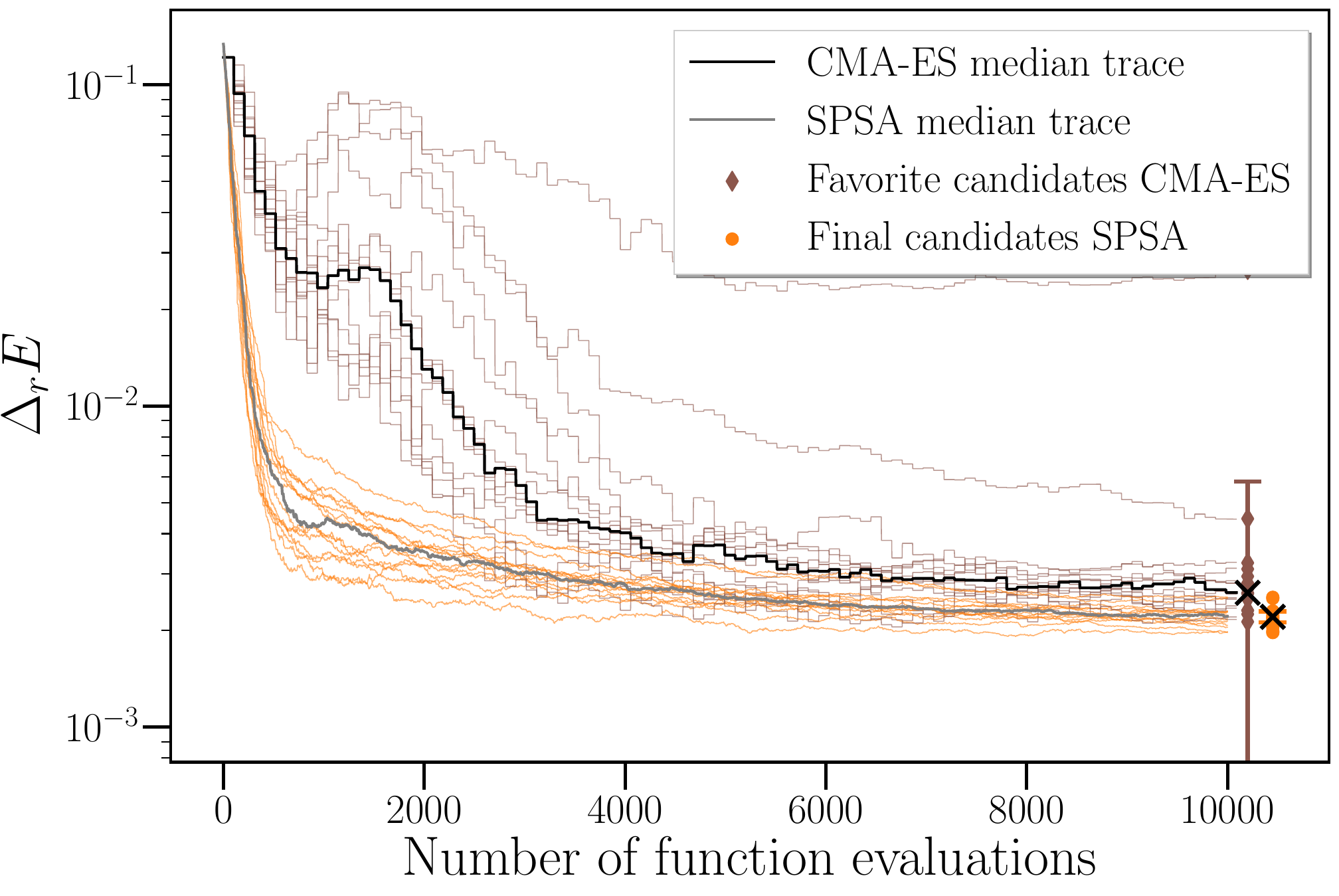}
    \label{fig:appendix_water_trace_onestage}
    \end{subfigure}
    
    \begin{subfigure}{0.5\textwidth}
    \centering
    \includegraphics[width=\textwidth]{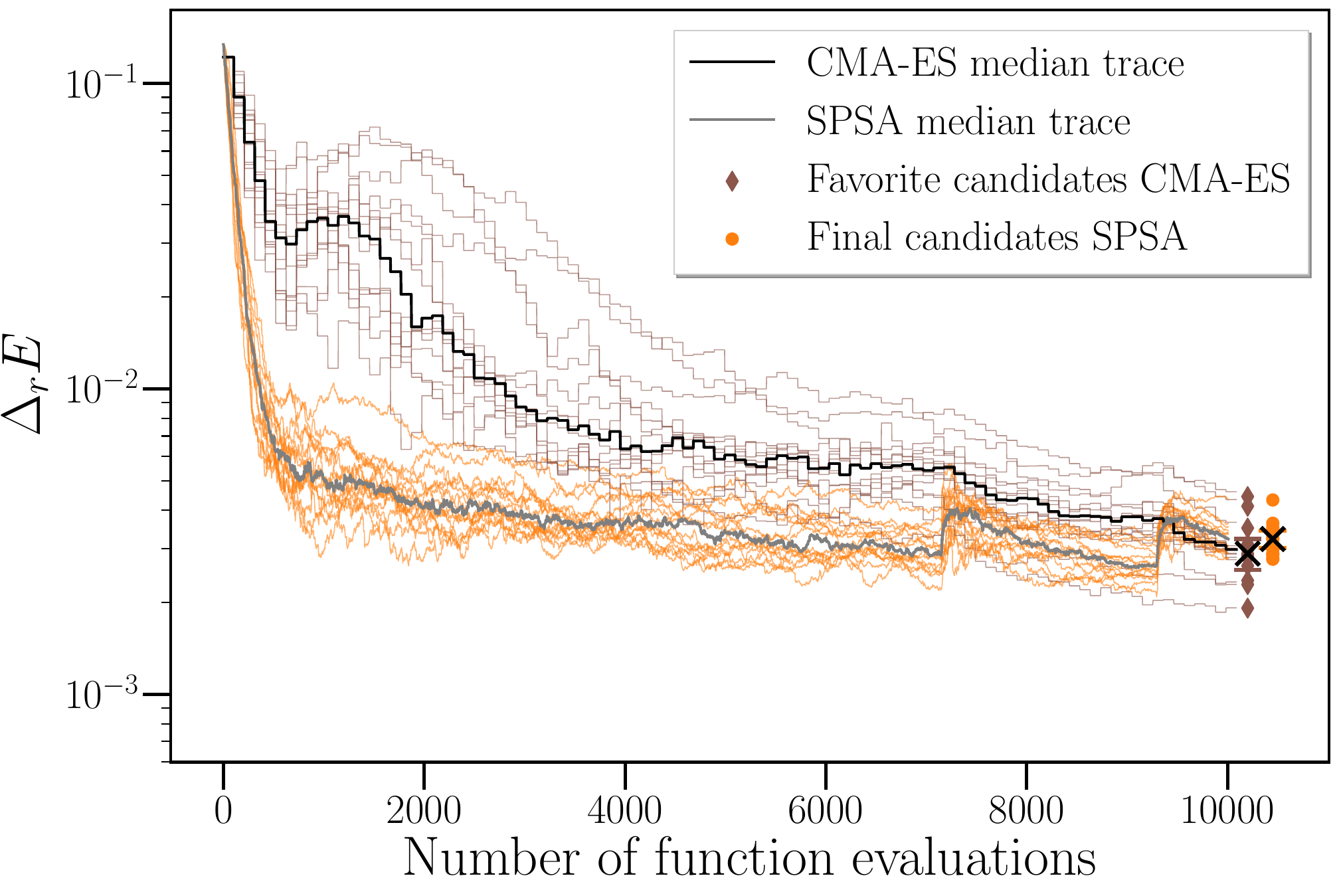}
    \label{fig:appendix_water_trace}
    \end{subfigure}

    \caption{Optimization traces of SPSA (orange and grey) and CMA-ES (blue and black) on the water molecule in the stretched configuration with \(10^8\) total shots.
    The thin lines depict an independent optimization trace, with the thick line on top being the median of the 15 independent runs.
    Top and bottom panel show the one- and three-stage sampling respectively.
    }
    \label{fig:appendix_water_trace_both}
\end{figure}

In this appendix, we provide examples of the convergence traces of SPSA and CMA-ES with their hyperparameter tuned.
We fix the total number of shots to \(10^8\), and separate the one- and three-stage sampling procedure.
In Fig.~\ref{fig:appendix_2x2_trace_both} we show the traces of the 2 x 2 Hubbard model. 
In Fig.~\ref{fig:appendix_2x3_trace_both} the traces of the 2 x 3 Hubbard model are depicted.
Finally, we show in Fig.\ref{fig:appendix_water_trace_both} the water molecule in stretched configuration optimization traces.
In all figures, the top panel shows the traces while using the one-stage sampling procedure.
The bottom panel shows the same traces but using the three-stage sampling method.

\clearpage

\end{document}